\documentclass{article}%
\usepackage{amsfonts}
\usepackage{amsmath}
\usepackage{amssymb}
\usepackage{graphicx}%
\newtheorem{theorem}{Theorem}

\newtheorem{definition}[theorem]{Definition}

\newtheorem{proposition}[theorem]{Proposition}
\newtheorem{remark}[theorem]{Remark}

\newenvironment{proof}[1][Proof]{\noindent\textbf{#1.} }{\ \rule{0.5em}{0.5em}}
\begin{document}

\title{Lagrangian Formalism for Multiform Fields on Minkowski Spacetime\thanks{This
paper is a new version of a paper published: \textit{Int. J. Theor. Phys}.
\textbf{40}, 2347-2378 (2001), which includes some additional references.}}
\author{A. M. Moya\thanks{Institute of Mathematics, Statistics and Scientific
Computation, IMECC-UNICAMP CP 6065, 13081-970 Campinas, SP, Brazil}, V. V.
Fern\'{a}ndez\thanks{Institute of Physics Gleb Wataghin, IFGW-UNICAMP,
13083-970 Campinas, SP, Brazil}, and W. A. Rodrigues, Jr.\thanks{Institute of
Mathematics, Statistics and Scientific Computation, IMECC-UNICAMP, CP 6065,
13081-970 Campinas, SP, Brazil and Center for Research and Technology,
CPTEC-UNISAL, Av. A. Garret 267, 13087-290 Campinas, SP, Brazil; e-mail:
walrod@ime.unicamp.br or walrod@cptec.br}}
\maketitle

\begin{abstract}
We present an introduction to the mathematical theory of the Lagrangian
formalism for multiform fields on Minkowski spacetime based on the multiform
and extensor calculus. Our formulation gives a unified mathematical
description for the main relativistic field theories including the
gravitational field (which however will be discussed in a separate paper). We
worked out several examples (including tricks of the trade), from simple to
very sophisticated ones (like, e.g., the Dirac-Hestenes field on the more
general gravitational background) which show the power and beauty of the formalism.

\end{abstract}

\begin{center}
{\LARGE \vspace*{0.3in}}
\end{center}

\section{Introduction}

It is now well known that the \emph{multiform } and \emph{extensor} fields
over Minkowski spacetime provides an unifying language for expressing the main
field equations of contemporany physics (Hestenes, 1966; Hestenes and Sobczyk,
1984; Rodrigues and de Souza, 1993, 1994; Moya 1999; Moya, Fern\'{a}ndez, and
Rodrigues, 2001), including gravitation\footnote{A complete review, including
many new mathematical topics of the overall subject has been published in a
special edition of \textit{Advances in Applied Clifford Algebras} (Moya,
Fern\'{a}ndez and Rodrigues, 2001). See also (Moya, Fern\'{a}ndez and
Rodrigues, 2003).}. However, a comprehensive \emph{multiform Lagrangian
formalism} using rigorous mathematics is still lacking, despite some previous
attempts (Lasenby, Doran and Gull, 1993; Rodrigues and de Souza, 1994;
Rodrigues et al,1995). In this paper we provides such a theory. In our
formalism several different kinds of Lagrangians which occurs in well known
physical theories are treated with the same mathematics. We include several
examples, from simple to sophisticated ones, which show the power of the
multiform calculus, together with the main identities (the tricks of trade)
necessary for the derivation of several equations of motion in the different
theories studied in this paper\footnote{In (Moya, Fern\'{a}ndez and Rodrigues,
2000a,b) we present a Lagrangian formulation for the gravitational field, as a
distortion field (an extensor field) on Minkowski spacetime. There, we show,
that the formulation of the same problem by (Lasenby, Doran and Gull,1998) is
non sequitur. A preliminary version of the gravitational field as a distortion
field appears in (Rodrigues and de Souza, 1993).}.

The power of the multiform calculus on Minkowski spacetime, $(M,\mathbf{\eta
},\tau_{\mathbf{\eta}},D^{\mathbf{\eta}})$\footnote{$M$ is a 4-dimensional
manifold oriented by $\tau_{\mathbf{\eta}}$ (the volume element 4-form) and
time oriented, which is diffeomorphic to $R^{4}$, $\mathbf{\eta}\in\sec
T_{2}^{0}(M)$ is a Lorentzian flat metric and $D^{\mathbf{\eta}}$ is the
Levi-Civita connection of $\mathbf{\eta}.$ For details, see (Sachs and
Wu,1977).} is enhanced once we use as a representation of this one an
\emph{affine space} ($M,\mathcal{M}^{\mathcal{*}}$), where $\mathcal{M}%
^{\mathcal{*}}$ the dual of $\mathcal{M(}\approx R^{4})$, is the vector space
of the structure. In the rest of this section we introduce some necessary
notations. Given a \textit{global coordinate system }over $M,$ say $M\ni
x\leftrightarrow x^{\mu}(x)\in R$ ($\mu=0,1,2,3 $) associated to a inertial
reference frame (Rodrigues and Rosa, 1989) at $x\in M$. $\left\langle \left.
\frac{\partial}{\partial x^{\mu}}\right|  _{x}\right\rangle $ and
$\left\langle \left.  dx^{\mu}\right|  _{x}\right\rangle $ are the
\textit{natural basis} for the tangent vector space $T_{x}M$ and the tangent
covector space $T_{x}^{*}M.$

We have,
\begin{align}
\mathbf{\eta}  &  =\eta_{\mu\nu}dx^{\mu}\otimes dx^{\nu},\nonumber\\
\eta_{\mu\nu}  &  =\mathbf{\eta}(\frac{\partial}{\partial x^{\mu}}%
,\frac{\partial}{\partial x^{\nu}})=diag(1,-1,-1,-1). \label{01}%
\end{align}

\begin{definition}
$T_{x}M\ni\mathbf{v}_{x}$ is said to be equipolent to $\mathbf{v}_{x^{\prime}%
}\in T_{x^{\prime}}M$ (written $\mathbf{v}_{x}=\mathbf{v}_{x^{\prime}}$) if
and only if
\begin{equation}
\mathbf{\eta}_{(x)}(\left.  \frac{\partial}{\partial x^{\mu}}\right|
_{x},\mathbf{v}_{x})=\mathbf{\eta}_{(x^{\prime})}(\left.  \frac{\partial
}{\partial x^{\mu}}\right|  _{x^{\prime}},\mathbf{v}_{x^{\prime}}),\text{
}(\mu=0,1,2,3).
\end{equation}

\end{definition}

Note that $\left.  \dfrac{\partial}{\partial x^{\beta}}\right|  _{x}=\left.
\dfrac{\partial}{\partial x^{\beta}}\right|  _{x^{\prime}}$ ($\beta=0,1,2,3$).

\begin{definition}
The set of equivalent classes of tangent vectors over the tangent bundle,
\begin{equation}
\mathcal{M}=\{\mathcal{C}_{\mathbf{v}_{x}}\mid\text{for all }x\in M\},
\label{04}%
\end{equation}
has a natural structure of vector space, it is called \textit{Minkowski vector
space}.
\end{definition}

Note that $\left\langle \mathcal{C}_{\left.  \tfrac{\partial}{\partial x^{\mu
}}\right|  _{x}}\right\rangle $ is a natural basis for $\mathcal{M}$
($\dim\mathcal{M}=4$). With the notations: $\vec{v}\equiv\mathcal{C}%
_{\mathbf{v}_{x}}$ and $\vec{e}_{\mu}\equiv\mathcal{C}_{\left.  \tfrac
{\partial}{\partial x^{\mu}}\right|  _{x}},$ we can write $\vec{v}=v^{\mu}%
\vec{e}_{\mu}.$

\begin{definition}
The 2-tensor over $\mathcal{M},$
\begin{equation}
\eta:\mathcal{M}\times\mathcal{M}\rightarrow R,
\end{equation}
such that for each $\vec{v}=\mathcal{C}_{\mathbf{v}_{x}}$ and $\vec
{w}=\mathcal{C}_{\mathbf{w}_{x}}\mathbf{\in}\mathcal{M}:\eta(\vec{v}%
\mathbf{,}$ $\vec{w}\mathbf{)=\eta}_{(x)}(\mathbf{v}_{x}\mathbf{,}$
$\mathbf{w}_{x}),$ for all $x\in M,$ is called \textit{Minkowski metric
tensor.}
\end{definition}

Note that, for each pair of basis vectors $\vec{e}_{\mu}\equiv\mathcal{C}%
_{\left.  \tfrac{\partial}{\partial x^{\mu}}\right|  _{x}}$ and $\vec{e}_{\nu
}\equiv\mathcal{C}_{\left.  \tfrac{\partial}{\partial x^{\nu}}\right|  _{x}},$
it holds
\begin{equation}
\eta(\vec{e}_{\mu},\vec{e}_{\nu})=diag(1,-1,-1,-1).
\end{equation}

\begin{definition}
The dual basis of $\left\langle \vec{e}_{\mu}\right\rangle $ will be
symbolized by $\left\langle \gamma^{\mu}\right\rangle ,$ i.e., $\gamma^{\mu
}\in\mathcal{M}^{*}\equiv\Lambda^{1}(\mathcal{M})$ and $\gamma^{\mu}(\vec
{e}_{\nu})=\delta_{\nu}^{\mu}.$
\end{definition}

To continue, we observe the existence of a fundamental isomorphism between
$\mathcal{M}$ and $\Lambda^{1}(\mathcal{M})$ given by,
\begin{equation}
\mathcal{M}\ni\vec{a}\mathbf{\leftrightarrow}a\in\Lambda^{1}(\mathcal{M}),
\label{07}%
\end{equation}
such that if $\vec{a}\mathbf{=}a^{\mu}\vec{e}_{\mu}$ then $a=\eta_{\mu\nu
}a^{\mu}\gamma^{\nu}$ and if $a=a_{\mu}\gamma^{\mu}$ then $\vec{a}%
\mathbf{=}\eta^{\mu\nu}a_{\mu}\vec{e}_{\nu}$, where $\eta_{\mu\nu}=\eta
(\vec{e}_{\mu},\vec{e}_{\nu})$, $\eta^{\mu\nu}=\eta_{\mu\nu}$.

\begin{remark}
To each basis vector $\vec{e}_{\mu}$ correspond a basis form $\gamma_{\mu
}=\eta_{\mu\nu}\gamma^{\nu}.$
\end{remark}

\begin{definition}
A scalar product of forms can be defined by
\begin{equation}
\Lambda^{1}(\mathcal{M})\times\Lambda^{1}(\mathcal{M})\ni(a,b)\mapsto a\cdot
b\in R, \label{07bis}%
\end{equation}
such that if $\vec{a}\mathbf{\leftrightarrow}a$ and $\vec{b}%
\mathbf{\leftrightarrow}b$ then $a\cdot b=\eta(\vec{a},\vec{b}).$
\end{definition}

\begin{remark}
$\gamma_{\mu}\cdot\gamma_{\nu}=\eta_{\mu\nu},$ $\gamma^{\mu}\cdot\gamma_{\nu
}=\delta_{\nu}^{\mu}$ ($\left\langle \gamma_{\mu}\right\rangle $ is called the
reciprocal basis of $\left\langle \gamma^{\mu}\right\rangle $) and
$\gamma^{\mu}\cdot\gamma^{\nu}=\eta^{\mu\nu}$. Thus, $\eta$ admits the
expansions $\eta=\eta_{\mu\nu}\gamma^{\mu}\otimes\gamma^{\nu}=\eta^{\mu\nu
}\gamma_{\mu}\otimes\gamma_{\nu}$.
\end{remark}

\begin{remark}
The oriented affine space $(M,\mathcal{M}^{\mathcal{*}})$ (oriented by
$\gamma^{5}=\gamma^{0}\wedge\gamma^{1}\wedge\gamma^{2}\wedge\gamma^{3} $) is a
representation of the Minkowski manifold.
\end{remark}

\begin{remark}
$(M,\mathcal{M}^{\mathcal{*}})$ equipped with the scalar product given by
eq.(\ref{07bis}) is a representation of Minkowski spacetime.
\end{remark}

\begin{definition}
Let $\langle x^{\mu}\rangle$ be a global affine coordinate system for
$(M,\mathcal{M}^{\mathcal{*}})$, relative to an arbitrary point $o$ $\in M$. A
position form associated to $x\in M$, is the form over $\mathcal{M}$ (designed
by the same letter), given by the following correspondence
\begin{equation}
M\ni x\leftrightarrows x=x^{\mu}\gamma_{\mu}\in\Lambda^{1}(\mathcal{M}).
\end{equation}

\end{definition}

\begin{remark}
We denote by $\mathcal{C\ell(M)\approx C\ell}(1,3)\approx\mathbf{H}(2)$, the
spacetime algebra, i.e., the Clifford algebra (Lounesto, 1997) of
$\mathcal{M}^{\mathcal{*}}$ equipped with the scalar product defined by
eq.(\ref{07bis}).
\end{remark}

\begin{remark}
As a vector space over the reals, we have $\mathcal{C\ell(M)=}\sum
\limits_{p=0}^{4}\Lambda^{p}(\mathcal{M}).$
\end{remark}

\begin{definition}
A smooth multiform field $A$ on Minkowski spacetime is a multiform valued
function of position form,
\begin{equation}
\Lambda^{1}(\mathcal{M})\ni x\mapsto A(x)\in\Lambda(\mathcal{M}) \label{010}%
\end{equation}

\end{definition}

\begin{definition}
Let $0\leq p,q\leq4$. A $(p,q)$-extensor $t$ is a linear mapping
\begin{equation}
t:\Lambda^{p}(\mathcal{M})\rightarrow\Lambda^{q}(\mathcal{M}). \label{011}%
\end{equation}

\end{definition}

\begin{remark}
The set of all $(p,q)$-extensors is denoted by $ext(\Lambda^{p}(\mathcal{M}%
),\Lambda^{q}(\mathcal{M})).$
\end{remark}

\begin{definition}
A smooth $(p,q)$-extensor field $t$ on Minkowski spacetime is a differentiable
$(p,q)$-extensor valued function of position form,
\begin{equation}
\Lambda^{1}(\mathcal{M})\ni x\mapsto t_{x}\in ext(\Lambda^{p}(\mathcal{M}%
),\Lambda^{q}(\mathcal{M})). \label{012}%
\end{equation}

\end{definition}

\begin{definition}
The $a$-directional derivative ($a$ is an arbitrary form) of a smooth
multiform field $X$, denoted as $a\cdot\partial X,$ is defined by
\begin{equation}
a\cdot\partial X=\underset{\lambda\rightarrow0}{\lim}\frac{X(x+\lambda
a)-X(x)}{\lambda}=\left.  \frac{d}{d\lambda}X(x+\lambda a)\right|
_{\lambda=0}. \label{013}%
\end{equation}

\end{definition}

\begin{remark}
The $\gamma_{\mu}$-directional derivative $\gamma_{\mu}\cdot\partial X$
coincides with the coordinate derivative $\frac{\partial X}{\partial x^{\mu}%
}.$ For short, we will use the notation $\partial_{\mu}\equiv\gamma_{\mu}%
\cdot\partial.$
\end{remark}

\begin{definition}
The gradient, divergence and curl of a smooth multiform field $X,$
respectively denoted by $\partial X,$ $\partial\lrcorner X$ and $\partial
\wedge X,$ are defined by
\begin{align}
\partial X  &  =\gamma^{\mu}(\partial_{\mu}X).\label{014a}\\
\partial\lrcorner X  &  =\gamma^{\mu}\lrcorner(\partial_{\mu}X).\label{014b}\\
\partial\wedge X  &  =\gamma^{\mu}\wedge(\partial_{\mu}X). \label{014c}%
\end{align}

\end{definition}

\begin{remark}
For any $X$, it holds $\partial X=\partial\lrcorner X+\partial\wedge X.$
\end{remark}

\section{Lagrangian mappings $(X,\partial*X)\mapsto\mathcal{L}(X,\partial*X)$}

Let $X$ be a smooth multiform field over the Minkowski spacetime $M$. Consider
the ordinary multiform\footnote{Called multivector derivatives in (Hestenes
and Sobczyk, 1984), where this concept, for the best of our knowledge, has
been introduced.} derivatives $\partial*X,$ where $*$ means any one of the
multiform products $(\lrcorner),(\wedge)$ or $($\textit{Clifford product}$)$
(i.e., the divergence, the curl or the gradient of $X$)\footnote{The dot
product of defintion 1.5 is extended in a natural way to all $Cl(\mathcal{M})$
as follows: $\forall X,Y\in\Lambda(\mathcal{M}),$ $X\cdot Y=\langle X\tilde
{Y}\rangle_{0}.$ Note that it is an operation different from the left and
right contractions. See (Rodrigues et al, 1995; Lounesto, 1999, Moya,
Fern\'{a}ndez and Rodrigues, 2000).}.

\begin{definition}
A differentiable scalar-valued function of two multiform variables,
\begin{equation}
\mathcal{L}:\operatorname{Im}X\times\operatorname{Im}\partial*X\rightarrow R,
\label{1.1}%
\end{equation}
(where Im$X$ means image of the multiform field $X$, etc.) will be called
Lagrangian mapping (\emph{LM}) associated to\ $X.$
\end{definition}

\begin{definition}
Let $X$ be any smooth multiform field, then $\widehat{\mathcal{L}}[X]$ is a
smooth scalar field defined by
\begin{equation}
\Lambda^{1}(\mathcal{M})\ni x\mapsto\widehat{\mathcal{L}}[X](x)\in R,
\label{1.2}%
\end{equation}
such that $\widehat{\mathcal{L}}[X](x)=\mathcal{L}[X(x),\partial*X(x)].$
\end{definition}

The operator $\widehat{\mathcal{L}}$ will be called Lagrangian operator and
the smooth scalar field $\widehat{\mathcal{L}}[X]$ will be called Lagrangian
scalar field associated to\textit{\ }$X.$

\begin{remark}
For abuse of notation, in what follows, the Lagrangian mapping $\mathcal{L}$
and the Lagrangian scalar field $\widehat{\mathcal{L}}[X]$ will be symbolized
simply by $(X,\partial*X)\mapsto\mathcal{L}(X,\partial*X).$
\end{remark}

\begin{definition}
To any LM $(X,\partial*X)\mapsto\mathcal{L}(X,\partial*X)$, the \emph{action}
for the multiform field $X$ (on $U\subseteq M$) is the scalar (i.e., a real
number)
\begin{equation}
S=\int_{U}\widehat{\mathcal{L}}[X](x)d^{4}x, \label{1.3N}%
\end{equation}
or, in abused notation $S=\int_{U}\mathcal{L}(X,\partial*X)$ $d^{4}x.$
\end{definition}

Take an arbitrary smooth multiform field $A,$ with the property
$A=\left\langle A\right\rangle _{X}$ (i.e., $A$ and $X$ are contains the same
grades), such that it vanishes on the boundary $\partial U$ (i.e., $\left.
A\right|  _{\partial U}=O$), and take an open set $S_{0}\subset R,$ containing zero.

\begin{definition}
The $\lambda$-parametrized smooth scalar field
\begin{equation}
\Lambda^{1}(\mathcal{M})\times S_{0}\ni(x,\lambda)\mapsto\widehat{\mathcal{L}%
}[X+\lambda A](x)\in R, \label{1.4}%
\end{equation}
will be called varied Lagrangian\textit{.}
\end{definition}

Note that $\widehat{\mathcal{L}}[X+\lambda A](x)=\mathcal{L}[X(x)+\lambda
A(x),\partial*X(x)+\lambda\partial*A(x)].$ Thus, for abuse of notation, the
varied Lagrangian would be sometimes denoted by $\mathcal{L}(X+\lambda
A,\partial*X+\lambda\partial*A).$

\begin{definition}
The $\lambda$-parametrized scalar (i.e., an ordinary scalar function of the
real variable $\lambda$)
\begin{equation}
S_{0}\ni\lambda\mapsto S(\lambda)\in R, \label{1.5}%
\end{equation}
such that
\begin{equation}
S(\lambda)=\int_{U}\widehat{\mathcal{L}}[X+\lambda A](x)d^{4}x,\text{ }
\label{1.6}%
\end{equation}
(or $S(\lambda)=\int_{U}\mathcal{L}(X+\lambda A,\partial*X+\lambda
\partial*A)d^{4}x$, in abused notation) is called varied action .
\end{definition}

\begin{definition}
Given any smooth multiform field $X$ and Lagrangian scalar field
$\widehat{\mathcal{L}}[X]$, the variational operator\textit{\ }$\delta_{X}$ is
defined by $\Lambda^{1}(\mathcal{M})\ni x\mapsto\delta_{X}\widehat
{\mathcal{L}}[X](x)\in R,$%
\begin{align}
\delta_{X}\widehat{\mathcal{L}}[X](x)  &  =\left.  \frac{d}{d\lambda}%
\widehat{\mathcal{L}}[X+\lambda A](x)\right|  _{\lambda=0}\nonumber\\
&  \left.  =\frac{d}{d\lambda}\mathcal{L}[X(x)+\lambda A(x),\partial
*X(x)+\lambda\partial*A(x)]\right|  _{\lambda=0}. \label{1.7a}%
\end{align}

\end{definition}

We simply write $\delta_{X}\mathcal{L}(X,\partial*X)=\frac{d}{d\lambda
}\mathcal{L}(X+\lambda A,\partial*X+\lambda\partial*A)|_{\lambda=0}$, in what follows.

In Lagrangian field theory, the dynamics of a multiform field $X$ is supposed
to satisfy the so-called \emph{condition of stationary action}\textit{,
}hereafter denoted\textit{\ }\emph{AP}\textit{, }i.e.\textit{,}
\begin{equation}
S^{\prime}(0)=0,\text{ for all }A\text{ such that }\left.  A\right|
_{\partial U}=O. \label{1.10}%
\end{equation}
The \emph{AP} can also be written as
\begin{equation}
\int_{U}\delta_{X}\mathcal{L}(X,\partial*X)d^{4}x=0,\text{ }\forall A\text{
}\mid\left.  A\right|  _{\partial U}=O,
\end{equation}

The \emph{AP} implies the so-called \emph{Euler-Lagrange equation (ELE)} for
the multiform field $X$\textit{\ (\emph{i.e., the field equation for} }%
$X$\textit{).}

\begin{proposition}
Given a dynamical variable $X$ on $U\subseteq M$, and a LM\emph{\ }%
$(X,\partial*X)\mapsto\mathcal{L}(X,\partial*X)$, where $*$ is respectively:
(a) $\lrcorner$, or (b) $\wedge$, or (c) the Clifford product, the AP implies
for the cases (a), (b), (c) respectively the following ELEs\emph{. }
\begin{align}
\text{(a)}  &  :\text{ }\partial_{X}\mathcal{L}(X,\partial\lrcorner
X)-\partial\wedge\partial_{\partial\lrcorner X}\mathcal{L}(X,\partial\lrcorner
X)=O.,\label{1.12}\\
\text{(b)}  &  :\text{ }\partial_{X}\mathcal{L}(X,\partial\wedge
X)-\partial\lrcorner\partial_{\partial\wedge X}\mathcal{L}(X,\partial\wedge
X)=O,\\
\text{(c)}  &  :\text{{}}\partial_{X}\mathcal{L}(X,\partial X)-\partial
\partial_{\partial X}\mathcal{L}(X,\partial X)=O.
\end{align}

\end{proposition}%

\proof

Here we prove only case (c), leaving the proofs of (a) and (b) to the reader.
The $X$-variation of $\mathcal{L}(X,\partial X)$ yields
\begin{equation}
\delta_{X}\mathcal{L}(X,\partial X)=A\cdot\partial_{X}\mathcal{L}(X,\partial
X)+\partial A\cdot\partial_{\partial X}\mathcal{L}(X,\partial X). \label{1.15}%
\end{equation}
Using the identity $(6.3)$, we have
\begin{equation}
\delta_{X}\mathcal{L}(X,\partial X)=A\cdot[\partial_{X}\mathcal{L}(X,\partial
X)-\partial\partial_{\partial X}\mathcal{L}(X,\partial X)]+\partial
\cdot[\partial_{a}(aA)\cdot\partial_{\partial X}\mathcal{L}(X,\partial X)],
\label{1.16}%
\end{equation}
and the \emph{AP} yields
\begin{equation}
\int_{U}A\cdot(\partial_{X}\mathcal{L}-\partial\partial_{\partial
X}\mathcal{L})\text{ }d^{4}x+\int_{U}\partial\cdot[\partial_{a}(aA)\cdot
\partial_{\partial X}\mathcal{L}]\text{ }d^{4}x=0, \label{1.17}%
\end{equation}
for all $A$ such that $\left.  A\right|  _{\partial U}=O$.

Using the Gauss-Stokes theorem with the boundary condition $\left.  A\right|
_{\partial U}=O$, the second term gives
\begin{align}
\int_{U}\partial\cdot[\partial_{a}(aA)\cdot\partial_{\partial X}%
\mathcal{L}]\text{ }d^{4}x  &  =\oint_{\partial U}\gamma^{\mu}\cdot
[\partial_{a}(aA)\cdot\partial_{\partial X}\mathcal{L}]\text{ }d^{3}S_{\mu
}\nonumber\\
&  =\oint_{\partial U}A\cdot(\gamma^{\mu}\partial_{\partial X}\mathcal{L}%
)\text{ }d^{3}S_{\mu}=0. \label{1.18}%
\end{align}

Putting eq.(\ref{1.18}) into eq.(\ref{1.17}) gives
\begin{equation}
\int_{U}A\cdot[\partial_{X}\mathcal{L}(X,\partial X)-\partial\partial
_{\partial X}\mathcal{L}(X,\partial X)]\text{ }d^{4}x=0,\text{ for all }A,
\label{1.19}%
\end{equation}

and due to the arbitrariness of $A,$ we get
\begin{equation}
\partial_{X}\mathcal{L}(X,\partial X)-\partial\partial_{\partial X}%
\mathcal{L}(X,\partial X)=O.
\endproof
\label{1.20}%
\end{equation}

\section{Lagrangian mappings $(X,\mathcal{D}*X)\mapsto\mathcal{L}%
(X,\mathcal{D}*X)$}

Let $X$ be a smooth multiform field on ($U\subseteq$ $M,\mathcal{M}%
^{\mathcal{*}}$) and let $h$ be an invertible $(1,1)$-extensor field.
($h_{x}:\mathcal{M}^{\mathcal{*}}\ni x\mapsto ext(\Lambda^{1}(\mathcal{M}%
),\Lambda^{1}(\mathcal{M}))$) called the \emph{gauge metric extensor field}
(which is representation of the gravitational field in the most general
possible gravitational theory over Minkowski spacetime) . Also, define
$h^{*}=(h^{-1})^{\dagger}=(h^{\dagger})^{-1}$. Consider the operators
$\mathcal{D}*$ where $*$ means any multiform product $(\lrcorner),(\wedge)$ or
the Clifford product acting on the set of smooth multivector fields. They are
called, the $h$-\emph{divergence} $\mathcal{D}\lrcorner X\equiv h^{\star
}(\partial_{a})\lrcorner\mathcal{D}_{a}X,$ the $h$-\emph{curl}
$\mathcal{D\wedge}X\equiv h^{\star}(\partial_{a})\wedge\mathcal{D}_{a}X$ and
the $h $-\emph{gradient} $\mathcal{D}X\equiv h^{\star}(\partial_{a}%
)\mathcal{D}_{a}X$.

$\mathcal{D}_{a}X$ here is a \emph{directional covariant derivative}, obtained
from the Levi-Civita directional covariant derivative $D_{a}$, $\mathcal{D}%
_{a}X=h(D_{a}h^{-1}(X))$ studied in the general geometrical algebraic theory
of connections developed in (Fern\'{a}ndez, Moya and Rodrigues, 2000b; Moya,
Fern\'{a}ndez and Rodrigues, 2000a,b)
\begin{equation}
\mathcal{D}_{a}X=a\cdot\partial X+\Omega(a)\times X.
\end{equation}
We briefly recall that, $\Omega$ is called second connection extensor field,
$\Omega_{x}:\Lambda^{1}(\mathcal{M})\rightarrow\Lambda^{2}(\mathcal{M})$,
$\forall x\in\mathcal{M}^{*}$, $\Omega_{x}(a)=-\frac{1}{2}\partial
_{n(x)}\wedge\mathcal{D}_{a}n(x)$.

In theories which make use of the gauge covariant derivative concept, the
action for the multiform field $X$ (on ($U\subseteq$ $M,\mathcal{M}%
^{\mathcal{*}}$)), with dynamics given by an a \emph{LM} $(X,\mathcal{D}%
*X)\mapsto\mathcal{L}(X,\mathcal{D}*X)$, the action is postulated to be the
scalar
\begin{equation}
S=\int_{U}\mathcal{L}(X,\mathcal{D}*X)d^{4}x. \label{2.3}%
\end{equation}

Take an arbitrary smooth multiform field $A$, with the property
$A=\left\langle A\right\rangle _{X}$, such that $\left.  A\right|  _{\partial
U}=O$ and take an open set $S_{0}\subset R,$ containing zero.

\begin{definition}
The $A$-varied action for the multivector field $X$ (on $U\subseteq M$) is the
$\lambda$-parametrized scalar
\begin{equation}
S(\lambda)=\int_{U}\mathcal{L}(X+\lambda A,\mathcal{D}*X+\lambda
\mathcal{D}*A)\text{ }d^{4}x. \label{2.4}%
\end{equation}

\end{definition}

The dynamics of the multiform field $X$ is supposed to satisfy the \emph{AP}
\[
S^{\prime}(0)=0,\text{ for all }A\text{ such that }\left.  A\right|
_{\partial U}=O.
\]
Equivalently, we have,
\begin{equation}
\int_{U}\delta_{X}\mathcal{L}(X,\mathcal{D}*X)\text{ }d^{4}x=0,\text{ for all
}A\text{ such that }\left.  A\right|  _{\partial U}=O, \label{2.5}%
\end{equation}
where $\delta_{X}\mathcal{L}(X,\mathcal{D}*X)=\left.  \tfrac{d}{d\lambda
}\mathcal{L}(X+\lambda A,\mathcal{D}*X+\lambda\mathcal{D}*A)\right|
_{\lambda=0}$ is the so-called $X$\textit{-variation of }$\mathcal{L}%
(X,\mathcal{D}*X)$.

\begin{proposition}
Given a dynamical variable $X$ and a LM $(X,\mathcal{D}*X)\mapsto$%
\linebreak$\mathcal{L}(X,\mathcal{D}*X)=\det(h)\ell(X,\mathcal{D}*X)$ where
$*$ is respectively: (a) $\lrcorner$, or (b) $\wedge$, or (c) the Clifford
product, the AP implies for the cases (a), (b), (c) the following ELEs\emph{,
}
\begin{align}
\text{(a)}  &  :\text{ }\partial_{X}\mathcal{\ell}(X,\mathcal{D}\lrcorner
X)-\mathcal{D}\wedge\partial_{\mathcal{D}\lrcorner X}\ell(X,\mathcal{D}%
\lrcorner X)=O,\label{2.6}\\
\text{(b)}  &  :\text{{}}\partial_{X}\mathcal{\ell}(X,\mathcal{D}\wedge
X)-\mathcal{D}\lrcorner\partial_{\mathcal{D}\wedge X}\mathcal{\ell
}(X,\mathcal{D}\wedge X)=O,\\
\text{(c)}  &  :\text{{}}\partial_{X}\mathcal{\ell}(X,\mathcal{D}%
X)-\mathcal{D}\partial_{\mathcal{D}X}\mathcal{\ell}(X,\mathcal{D}X)=O.
\end{align}

\end{proposition}%

\proof

We prove only the case (b).Using the multiform identity ( 6.8) the
$X$-variation of $\mathcal{L}(X,\mathcal{D}\wedge X)$ yields
\begin{align}
\delta_{X}\mathcal{L}(X,\mathcal{D}\wedge X)  &  =\det(h)[A\cdot\partial
_{X}\mathcal{\ell}(X,\mathcal{D}\wedge X)+\mathcal{D}\wedge A\cdot
\partial_{\mathcal{D}\wedge X}\mathcal{\ell}(X,\mathcal{D}\wedge
X)]\nonumber\\
&  =\det(h)A\cdot[\partial_{X}\mathcal{\ell}(X,\mathcal{D}\wedge
X)-\mathcal{D}\lrcorner\partial_{\mathcal{D}\wedge X}\mathcal{\ell
}(X,\mathcal{D}\wedge X)]+\nonumber\\
&  =\partial\cdot[\det(h)\partial_{a}(h^{\star}(a)\wedge A)\cdot
\partial_{\mathcal{D}\wedge X}\mathcal{\ell}(X,\mathcal{D}\wedge X)].
\label{2.10}%
\end{align}

The \emph{AP} action yields,
\begin{equation}
\int_{U}\det(h)A\cdot(\partial_{X}\mathcal{\ell}-\mathcal{D}\lrcorner
\partial_{\mathcal{D}\wedge X}\mathcal{\ell})\text{ }d^{4}x+\int_{U}%
\partial\cdot[\det(h)\partial_{a}(h^{\star}(a)\wedge A)\cdot\partial
_{\mathcal{D}\wedge X}\mathcal{\ell}]\text{ }d^{4}x=0, \label{2.11}%
\end{equation}
for all $A$ such that $\left.  A\right|  _{\partial U}=O$.

Using the Gauss-Stokes theorem with the boundary condition $\left.  A\right|
_{\partial U}=O,$ the second term gives,
\begin{align}
&  \int_{U}\partial\cdot[\det(h)\partial_{a}(h^{\star}(a)\wedge A)\cdot
\partial_{\mathcal{D}\wedge X}\mathcal{\ell}]\text{ }d^{4}x\nonumber\\
&  =\oint_{\partial U}\det(h)\gamma^{\mu}\cdot[\partial_{a}(h^{\star}(a)\wedge
A)\cdot\partial_{\mathcal{D}\wedge X}\mathcal{\ell}]\text{ }d^{3}S_{\mu
}\nonumber\\
&  =\oint_{\partial U}\det(h)A\cdot[h^{\star}(\gamma^{\mu})\lrcorner
\partial_{\mathcal{D}\wedge X}\mathcal{\ell}]\text{ }d^{3}S_{\mu}=0.
\label{2.12}%
\end{align}

Putting eq.(\ref{2.12}) into eq.(\ref{2.11})$,$ we have
\[
\int_{U}\det(h)A\cdot[\partial_{X}\mathcal{\ell}(X,\mathcal{D}\wedge
X)-\mathcal{D}\lrcorner\partial_{\mathcal{D}\wedge X}\mathcal{\ell
}(X,\mathcal{D}\wedge X)]\text{ }d^{4}x=0,\text{ for all }A.
\]

and due to the arbitrariness of $A,$ we finally get
\begin{equation}
\partial_{X}\mathcal{\ell}(X,\mathcal{D}\wedge X)-\mathcal{D}\lrcorner
\partial_{\mathcal{D}\wedge X}\mathcal{\ell}(X,\mathcal{D}\wedge X)=O.
\endproof
\label{2.13}%
\end{equation}

\begin{remark}
The proofs of (a) and (b) can be easily obtained by using the multiform
identities ( 6.7) and (6.9).
\end{remark}

\section{Lagrangian Mapping $(\psi,\mathcal{D}^{s}\psi)\mapsto\mathcal{L}%
(\psi,\mathcal{D}^{s}\psi)$}

Let $\psi$ be a smooth Dirac-Hestenes spinor field (\emph{DHSF}) on
($U\subseteq M,\mathcal{M}^{\mathcal{*}}$) We can take the \emph{gauge spinor
derivative} (Rodrigues et al,1995; Fern\'{a}ndez, Moya and Rodrigues, 2000b)
$\mathcal{D}^{s}\psi\equiv h^{\star}(\partial_{a})\mathcal{D}_{a}^{s}\psi$
(recall that $\mathcal{D}_{a}^{s}\psi\equiv a\cdot\partial\psi+\frac{1}%
{2}\Omega(a)\psi$ is the \emph{directional spinor derivative}) and consider a
\emph{LM\ }$(\psi,\mathcal{D}^{s}\psi)\mapsto\mathcal{L}(\psi,\mathcal{D}%
^{s}\psi).$

\begin{definition}
The action for a DHSF $\psi$ (on $U\subseteq M$) is
\begin{equation}
S=\int_{U}\mathcal{L}(\psi,\mathcal{D}^{s}\psi)\text{ }d^{4}x. \label{3.1}%
\end{equation}

\end{definition}

If we take an arbitrary smooth \emph{DHSF} $\eta,$ such that $\left.
\eta\right|  _{\partial U}=O$, then, the so-called $\psi$-variation of
$\mathcal{L}(\psi,\mathcal{D}^{s}\psi)$ is%

\begin{equation}
\delta_{\psi}\mathcal{L}(\psi,\mathcal{D}^{s}\psi)=\left.  \tfrac{d}{d\lambda
}\mathcal{L}(\psi+\lambda\eta,\mathcal{D}^{s}\psi+\lambda\mathcal{D}^{s}%
\eta)\right|  _{\lambda=0}. \label{3.2}%
\end{equation}

\begin{proposition}
Given a DHSF $\psi,$ as dynamical variable, and a LM\emph{\ }$(\psi
,\mathcal{D}^{s}\psi)\mapsto\mathcal{L}(\psi,\mathcal{D}^{s}\psi)$ the AP
\[
\int_{U}\delta\psi\mathcal{L}(\psi,\mathcal{D}^{s}\psi)d^{4}x=0,
\]
for all $\eta$ such that $\left.  \eta\right|  _{\partial U}=O$, implies the
ELE,
\begin{equation}
\partial_{\psi}\mathcal{\ell}(\psi,\mathcal{D}^{s}\psi)-\mathcal{D}%
^{s}\partial_{\mathcal{D}^{s}\psi}\ell(\psi,\mathcal{D}^{s}\psi)=O.
\label{3.5}%
\end{equation}

\end{proposition}%

\proof

The $\psi$-variation of $\mathcal{L}(\psi,\mathcal{D}^{s}\psi)$ is
\begin{equation}
\delta\psi\mathcal{L}(\psi,\mathcal{D}^{s}\psi)=\det(h)[\eta\cdot
\partial_{\psi}\mathcal{\ell}(\psi,\mathcal{D}^{s}\psi)+\mathcal{D}^{s}%
\eta\cdot\partial_{\mathcal{D}^{s}\psi}\ell(\psi,\mathcal{D}^{s}\psi)],
\label{3.6}%
\end{equation}
and using the multiform identity (6.14) we have
\begin{align}
\delta\psi\mathcal{L}(\psi,\mathcal{D}^{s}\psi)  &  =\det(h)\eta\cdot
[\partial_{\psi}\mathcal{\ell}(\psi,\mathcal{D}^{s}\psi)-\mathcal{D}%
^{s}\partial_{\mathcal{D}^{s}\psi}\ell(\psi,\mathcal{D}^{s}\psi)]\nonumber\\
&  +\partial\cdot[\det(h)\partial_{a}(h^{\star}(a)\eta)\cdot\partial
_{\mathcal{D}^{s}\psi}\ell(\psi,\mathcal{D}^{s}\psi)]. \label{3.7}%
\end{align}

Thus, the \emph{AP} can be written,
\begin{equation}
\int_{U}\det(h)\eta\cdot(\partial_{\psi}\mathcal{\ell}-\mathcal{D}^{s}%
\partial_{\mathcal{D}^{s}\psi}\ell)\text{ }d^{4}x+\int_{U}\partial\cdot
[\det(h)\partial_{a}(h^{\star}(a)\eta)\cdot\partial_{\mathcal{D}^{s}\psi}%
\ell]\text{ }d^{4}x=0, \label{3.8}%
\end{equation}
for all $\eta$ such that $\left.  \eta\right|  _{\partial U}=O.$

The second term can be integrated using the Gauss-Stokes theorem with the
boundary condition $\left.  \eta\right|  _{\partial U}=O$. We get,
\begin{align}
\int_{U}\partial\cdot[\det(h)\partial_{a}(h^{\star}(a)\eta)\cdot
\partial_{\mathcal{D}^{s}\psi}\ell]\text{ }d^{4}x  &  =\oint_{\partial
U}\gamma^{\mu}\cdot[\det(h)\partial_{a}(h^{\star}(a)\eta)\cdot\partial
_{\mathcal{D}^{s}\psi}\ell]\text{ }d^{3}S_{\mu}\nonumber\\
&  =\oint_{\partial U}\det(h)\eta\cdot[h^{\star}(\gamma^{\mu})\partial
_{\mathcal{D}^{s}\psi}\ell]\text{ }d^{3}S_{\mu}=0.\nonumber\\
&  \label{3.9}%
\end{align}

Putting eq.(\ref{3.9}) into eq.(\ref{3.8}), we get
\begin{equation}
\int_{U}\det(h)\eta\cdot[\partial_{\psi}\mathcal{\ell}(\psi,\mathcal{D}%
^{s}\psi)-\mathcal{D}^{s}\partial_{\mathcal{D}^{s}\psi}\ell(\psi
,\mathcal{D}^{s}\psi)]\text{ }d^{4}x=0,\text{ for all }\eta. \label{3.10}%
\end{equation}

And since $\eta$ is arbitrary, it follows that
\begin{equation}
\partial_{\psi}\mathcal{\ell}(\psi,\mathcal{D}^{s}\psi)-\mathcal{D}%
^{s}\partial_{\mathcal{D}^{s}\psi}\ell(\psi,\mathcal{D}^{s}\psi)=O.
\endproof
\label{3.11}%
\end{equation}

\section{Examples}

\subsection{Maxwell and Dirac-Hestenes Lagrangians on Minkowski spacetime}

(a) The Lagrangian associated to the Maxwell field $A:\mathcal{M}%
^{\mathcal{*}}\rightarrow\Lambda^{1}(\mathcal{M})$ (i.e, the electromagnetic
potential) generated by an electric charge current density $J:\mathcal{M}%
^{\mathcal{*}}\rightarrow\Lambda^{1}(\mathcal{M}),$ is
\begin{equation}
\mathcal{L}(A,\partial\wedge A)=-\tfrac{1}{2\mu_{0}}(\partial\wedge
A)\cdot(\partial\wedge A)-A\cdot J. \label{4.1}%
\end{equation}

The Euler-Lagrange is then, according to previous results
\begin{equation}
\partial_{A}\mathcal{L}(A,\partial\wedge A)-\partial\lrcorner\partial
_{\partial\wedge A}\mathcal{L}(A,\partial\wedge A)=O. \label{4.2}%
\end{equation}

Then, the Maxwell field $A$ and the Faraday field $F=\partial\wedge A,$
satisfy the equations
\begin{equation}
\partial\lrcorner(\partial\wedge A)=\mu_{0}J,\text{ }\partial F=\mu_{0}J.
\label{4.4}%
\end{equation}
The second equation in (\ref{4.4}) is \emph{Maxwell equation} in the spacetime
calculus formalism (Hestenes,1966).

(b) In quantum mechanics, the Lagrangian\footnote{A thoughtful study of the
Dirac-Hestenes Lagrangian which shows hidden assumptions in usual presentation
can be found in (De Leo et al, 1999)} associated to the \emph{DHSF}%
\footnote{\emph{DHSF} are certain equivalence classes of even sections of
$\mathcal{C\ell(M)}$. For details, see (Rodrigues et al, 1995).}
$\psi:\mathcal{M}^{\mathcal{*}}\rightarrow\Lambda^{0}(\mathcal{M})+\Lambda
^{2}(\mathcal{M})+\Lambda^{4}(\mathcal{M}) $ , corresponding to a particle
with mass $m,$ electric charge $e$ and spin $\frac{1}{2}$ (i.e., a Dirac
particle) in interaction with the Maxwell field $A$ , is
\begin{equation}
\mathcal{L}(\psi,\partial\psi)=\hslash(\partial\psi\mathbf{i}\gamma_{3}%
)\cdot\psi-e(A\psi\gamma_{0})\cdot\psi-mc\psi\cdot\psi, \label{4.6}%
\end{equation}
where $\mathbf{i}=\gamma_{0}\gamma_{1}\gamma_{2}\gamma_{3}$. To get the
\emph{ELE} we need%

\begin{align}
\partial_{\psi}\mathcal{L}(\psi,\partial\psi)  &  =\hslash\partial
\psi\mathbf{i}\gamma_{3}-2eA\psi\gamma_{0}-2mc\psi,\nonumber\\
\partial_{\partial\psi}\mathcal{L}(\psi,\partial\psi)  &  =-\hslash
\partial_{\partial\psi}(\partial\psi\cdot\psi\mathbf{i}\gamma_{3}%
)=-\hslash\psi\mathbf{i}\gamma_{3}. \label{4.8}%
\end{align}
where the following multiform derivative formulas have been used,
\begin{equation}
\partial_{X}(X\cdot X)=2X,\text{ }\partial_{X}(X\cdot Y)=\left\langle
Y\right\rangle _{X},\text{ }\partial_{X}[(YXZ)\cdot X]=\left\langle
YXZ+\widetilde{Y}X\widetilde{Z}\right\rangle _{X}. \label{4.9}%
\end{equation}

Thus, the \emph{DHSF} $\psi$ satisfies
\begin{equation}
\hslash\partial\psi\mathbf{i}\sigma_{3}-eA\psi=mc\psi\gamma_{0},\text{ }%
\sigma_{3}=\gamma_{3}\gamma_{0} \label{4.9n}%
\end{equation}
which is the expression of the Dirac equation (called the Dirac-Hestenes
equation (Hestenes, 1996)) in the spacetime calculus formalism.\vspace*{0.3in}

\subsection{\textit{\ }Lagrangian for Maxwell and Dirac-Hestenes fields on a
gravitational field background}

In the flat spacetime formulation of the most general possible gravitational
field theory (which includes curvature and torsion), this field is described
by an invertible (1,1)- extensor field $h$ (section 3).

(a) The \emph{dynamics} of a Maxwell field $A$ generated by a electric charge
current density $J$, moving in a background gravitational field is postulated
to derived from the \emph{AP} and the following Lagrangian
\begin{equation}
\mathcal{L}(A,\mathcal{D}\wedge A)=\det(h)[-\tfrac{1}{2\mu_{0}}(\mathcal{D}%
\wedge A)\cdot(\mathcal{D}\wedge A)-A\cdot J]. \label{4.10}%
\end{equation}

Then (using the identities in remark 4 in the Appendix), we obtain that $A$
and $F=\mathcal{D}\wedge A$ satisfy
\begin{equation}
\mathcal{D}\lrcorner(\mathcal{D}\wedge A)=\mu_{0}J,\text{ }\mathcal{D}%
F=\mu_{0}J. \label{4.13}%
\end{equation}

(b) The dynamics of a \emph{DHSF} $\psi$ corresponding to a particle with mass
$m,$ electric charge $e$ and spin $\frac{1}{2}$ (i.e., a Dirac particle), is
supposed to be governed by the \emph{AP} with Lagrangian
\begin{equation}
\mathcal{L}(\psi,\mathcal{D}^{s}\psi)=\det(h)[\hslash(\mathcal{D}^{s}%
\psi\mathbf{i}\gamma_{3})\cdot\psi-e(A\psi\gamma_{0})\cdot\psi-mc\psi\cdot
\psi], \label{4.14}%
\end{equation}

Then (using the identities mentioned in remark 5 of Appendix) we get that the
\emph{DHSF} $\psi$ satisfies
\begin{equation}
\hslash\mathcal{D}^{s}\psi\mathbf{i}\sigma_{3}-eA\psi=mc\psi\gamma_{0}.
\label{4.18}%
\end{equation}

\section{ Appendix: fundamental identities used in Lagrangian formalism}

\begin{proposition}
For all smooth multiform fields $X,Y$ and 1-form field $a,$ it holds
\begin{align}
(\partial\lrcorner X)\cdot Y+X\cdot(\partial\wedge Y)  &  =\partial
\cdot[\partial_{a}(a\lrcorner X)\cdot Y].\label{1.a}\\
(\partial\wedge X)\cdot Y+X\cdot(\partial\lrcorner Y)  &  =\partial
\cdot[\partial_{a}(a\wedge X)\cdot Y].\label{1.b}\\
(\partial X)\cdot Y+X\cdot(\partial Y)  &  =\partial\cdot[\partial
_{a}(aX)\cdot Y]. \label{1.c}%
\end{align}

\end{proposition}%

\proof

In order to prove the first identity (\ref{1.a}), we use the definitions of
divergence and curl of a multiform field and the algebraic identity
$(a\lrcorner B)\cdot C=B\cdot(a\wedge C),$ where $a$ is a 1-fom and $B,C$ are
multiforms. Then,
\begin{align}
(\partial\lrcorner X)\cdot Y+X\cdot(\partial\wedge Y)  &  =(\gamma^{\mu
}\lrcorner\gamma_{\mu}\cdot\partial X)\cdot Y+X\cdot(\gamma^{\mu}\wedge
\gamma_{\mu}\cdot\partial Y)\nonumber\\
&  =\gamma_{\mu}\cdot\partial(\gamma^{\mu}\lrcorner X)\cdot Y+(\gamma^{\mu
}\lrcorner X)\cdot(\gamma_{\mu}\cdot\partial Y)\nonumber\\
&  =\gamma_{\mu}\cdot\partial(\gamma^{\mu}\lrcorner X)\cdot Y, \label{a2}%
\end{align}
but, it is not difficult to transform the right side of (\ref{a2}) into a
divergence of a 1-form field. Indeed,
\begin{align}
\gamma_{\mu}\cdot\partial(\gamma^{\mu}\lrcorner X)\cdot Y  &  =\gamma_{\beta
}\cdot\partial[\gamma^{\beta}\cdot\gamma_{\mu}(\gamma^{\mu}\lrcorner X)\cdot
Y]\nonumber\\
&  =\gamma^{\beta}\cdot\gamma_{\beta}\cdot\partial[\gamma_{\mu}(\gamma^{\mu
}\lrcorner X)\cdot Y]\nonumber\\
&  =\partial\cdot[\partial_{a}(a\lrcorner X)\cdot Y]. \label{a3}%
\end{align}

Putting (\ref{a2}) into (\ref{a3}), complete the proof.%
\endproof

\begin{remark}
These identities are necessary, e.g., in the derivation from the AP of the ELE
equations for a multiform field $X$ with dynamics governed by LM
$(X,\partial\lrcorner X)\mapsto\mathcal{L}(X,\partial\lrcorner X)$,
$(X,\partial\wedge X)\mapsto\mathcal{L}(X,\partial\wedge X)$ or $(X,\partial
X)\mapsto\mathcal{L}(X,\partial X)$.
\end{remark}

The identity (\ref{1.b}) can be proved by using (\ref{1.a}) and, once again,
the algebraic identity $(a\lrcorner B)\cdot C=B\cdot(a\wedge C)$. We have,
\begin{align}
(\partial\wedge X)\cdot Y+X\cdot(\partial\lrcorner Y)  &  =(\partial\lrcorner
Y)\cdot X+Y\cdot(\partial\wedge X)=\partial\cdot[\partial_{a}(a\lrcorner
Y)\cdot X]\nonumber\\
&  =\partial\cdot[\partial_{a}Y\cdot(a\wedge X)]=\partial\cdot[\partial
_{a}(a\wedge X)\cdot Y]. \label{a.4}%
\end{align}

Identity (\ref{1.c}) can be proved without difficulties by adding (\ref{1.a})
and (\ref{1.b}),

\begin{proposition}
For all smooth multiform fields $X,Y$ and 1-form field $a,$ it holds
\begin{align}
(\mathcal{D}\lrcorner X)\cdot Y+X\cdot(\mathcal{D}\wedge Y)  &  =\det
(h^{-1})\partial\cdot[\det(h)\partial_{a}(h^{\star}(a)\lrcorner X)\cdot
Y].\label{2.a}\\
(\mathcal{D}\wedge X)\cdot Y+X\cdot(\mathcal{D}\lrcorner Y)  &  =\det
(h^{-1})\partial\cdot[\det(h)\partial_{a}(h^{\star}(a)\wedge X)\cdot
Y].\label{2.b}\\
(\mathcal{D}X)\cdot Y+X\cdot(\mathcal{D}Y)  &  =\det(h^{-1})\partial\cdot
[\det(h)\partial_{a}(h^{\star}(a)X)\cdot Y]. \label{2.c}%
\end{align}

\end{proposition}%

\proof

To prove the identity (\ref{2.a}), we shall need to use two important
multiform identities relating the gauge covariant divergence and the ordinary
divergence, and the gauge covariant curl and the ordinary curl, respectively
(Fern\'{a}ndez, Moya and Rodrigues, 2000b)
\begin{equation}
\mathcal{D}\lrcorner A=\det(h^{-1})\underline{h}[\partial\lrcorner
\det(h)\underline{h}^{-1}(A)],\text{ }\mathcal{D}\wedge A=\underline{h}%
^{\star}[\partial\wedge\underline{h}^{\dagger}(A)]. \label{a6}%
\end{equation}
and the multiform identity (\ref{1.a}) above. We have
\begin{align}
(\mathcal{D}\lrcorner X)\cdot Y+X\cdot(\mathcal{D}\wedge Y)  &  =\det
(h^{-1})\underline{h}[\partial\lrcorner\det(h)\underline{h}^{-1}(X)]\cdot
Y+X\cdot\underline{h}^{\star}[\partial\wedge\underline{h}^{\dagger
}(Y)]\nonumber\\
&  =\det(h^{-1})\{[\partial\lrcorner\det(h)\underline{h}^{-1}(X)]\cdot
\underline{h}^{\dagger}(Y)\nonumber\\
&  +\det(h)\underline{h}^{-1}(X)\cdot[\partial\wedge\underline{h}^{\dagger
}(Y)]\}\nonumber\\
&  =\det(h^{-1})\partial\cdot[\partial_{a}(a\lrcorner\det(h)\underline{h}%
^{-1}(X))\cdot\underline{h}^{\dagger}(Y)]\nonumber\\
&  =\det(h^{-1})\partial\cdot[\det(h)\partial_{a}(a\lrcorner\underline{h}%
^{-1}(X))\cdot\underline{h}^{\dagger}(Y)]. \label{a7}%
\end{align}
Using the algebraic identity $a\lrcorner\underline{t}(B)=\underline
{t}[\underline{t}^{\dagger}(a)\lrcorner B],$ where $\underline{t}$ is the
extension\footnote{Given a (1,1)-extensor $t$ over $\Lambda(\mathcal{M})$ its
extension $\underline{t}$ , a general extensor over $\Lambda(\mathcal{M})$, is
defined by $\underline{t}(X)=1.X+\sum_{k=1}^{n}\frac{1}{k!}t(\gamma^{j_{1}%
})\wedge...\wedge t(\gamma^{j_{k}})(\gamma_{j_{1}}\wedge...\wedge\gamma
_{j_{k}})\cdot X$} of a (1,1) extensor $t$, $a$ is a 1-form and $B$ is a
multiform, we can write
\begin{equation}
\lbrack a\lrcorner\underline{h}^{-1}(X)]\cdot\underline{h}^{\dagger
}(Y)=\underline{h}^{-1}[\underline{h}^{\star}(a)\lrcorner X]\cdot\underline
{h}^{\dagger}(Y)=\underline{h}\underline{h}^{-1}[\underline{h}^{\star
}(a)\lrcorner X]\cdot Y=[\underline{h}^{\star}(a)\lrcorner X]\cdot Y.
\label{A8}%
\end{equation}

Putting (\ref{A8}) into the right side of (\ref{a7}), completes the proof.

The second identity (\ref{2.b}) can be proved using (\ref{2.a}), the algebraic
identity $(a\lrcorner B)\cdot C=B\cdot(a\wedge C),$ and following analogous
steps just used in demonstrating the identity (\ref{1.b}). The third identity
(\ref{2.c}) can be proved easily by adding (\ref{2.a}) and (\ref{2.b}).%
\endproof

\begin{remark}
These multiform identities are necessary in order to derive from the
AP\emph{\ }the ELEs for multiform fields $X$ with Lagrangian mappings
$(X,\mathcal{D}\lrcorner X)\mapsto\mathcal{L}(X,\mathcal{D}\lrcorner X)$, or
$(X,\mathcal{D}\wedge X)\mapsto\mathcal{L}(X,\mathcal{D}\wedge X)$ or
$(X,\mathcal{D}X)\mapsto\mathcal{L}(X,\mathcal{D}X)$.

\begin{proposition}
For all smooth \emph{DHSF} $\psi$ and $\varphi,$ it holds
\begin{align}
(\mathcal{D}^{s}\psi)\cdot\varphi+\psi\cdot(\mathcal{D}^{s}\varphi)  &
=(\mathcal{D}\psi)\cdot\varphi+\psi\cdot(\mathcal{D}\varphi).\label{A9N}\\
(\mathcal{D}^{s}\psi)\cdot\varphi+\psi\cdot(\mathcal{D}^{s}\varphi)  &
=\det(h^{-1})\partial\cdot\lbrack\det(h)\partial_{a}(h^{\star}(a)\psi
)\cdot\varphi]. \label{a.9}%
\end{align}

\end{proposition}
\end{remark}

To prove (\ref{A9N}), note that for a smooth spinor field $\psi,$ we have
\begin{equation}
\mathcal{D}\psi=\mathcal{D}^{s}\psi-\tfrac{1}{2}h^{\star}(\partial_{a}%
)\psi\Omega(a). \label{a.11}%
\end{equation}

Thus, by using (\ref{a.11}) with the \emph{DHSF} $\psi$ and $\varphi,$ we can
write
\begin{align}
(\mathcal{D}^{s}\psi)\cdot\varphi+\psi\cdot(\mathcal{D}^{s}\varphi)  &
=[\mathcal{D}\psi+\tfrac{1}{2}h^{\star}(\partial_{a})\psi\Omega(a)]\cdot
\varphi+\psi\cdot[\mathcal{D}\varphi+\tfrac{1}{2}h^{\star}(\partial
_{a})\varphi\Omega(a)]\nonumber\\
&  =(\mathcal{D}\psi)\cdot\varphi+\psi\cdot(\mathcal{D}\varphi)\nonumber\\
&  -\tfrac{1}{2}h^{\star}(\partial_{a})\cdot\varphi\Omega(a)\widetilde{\psi
}-\tfrac{1}{2}h^{\star}(\partial_{a})\cdot\psi\Omega(a)\widetilde{\varphi},
\label{A.12}%
\end{align}
but, the last two terms yield zero, as can be seen from,
\begin{align}
\tfrac{1}{2}h^{\star}(\partial_{a})\cdot[\varphi\Omega(a)\widetilde{\psi}%
+\psi\Omega(a)\widetilde{\varphi}]  &  =\tfrac{1}{2}h^{\star}(\partial
_{a})\cdot[\varphi\Omega(a)\widetilde{\psi}-(\varphi\Omega(a)\widetilde{\psi
})^{\sim}]\nonumber\\
&  =\tfrac{1}{2}h^{\star}(\partial_{a})\cdot2\left\langle \varphi
\Omega(a)\widetilde{\psi}\right\rangle _{2}=0, \label{A.13}%
\end{align}
which completes the proof. Eq.(\ref{a.9}) follows directly from (\ref{A9N})
and (6.9).%
\endproof
\medskip

\textbf{References\medskip}

{\footnotesize De Leo, S., Oziewicz, Z., Rodrigues, W. A. , Jr., and Vaz, J.,
Jr. (1999). The Dirac-Hestenes Lagrangian, \emph{Int. J. Theor. Phys}.
\textbf{38}, 2347-2367.}

{\footnotesize Moya, A. M. (1999). Lagrangian formalism for Multivector fields
on Minkowski spacetime, Ph.D. Thesis, IMECC-UNICAMP.}

{\footnotesize Moya, A. M., Fern\'{a}ndez, V. V., and Rodrigues, W. A., Jr.
(2001). Multivector and Extensor Calculus, \ \textit{Advances in Applied
Clifford Algebras }\textbf{11}(S3), 1-103, (2001)}

{\footnotesize Moya, A. M., Fern\'{a}ndez, V. V., and Rodrigues, W. A., Jr.
(2003). Gravitational Fields as Distortion Fields on Minkowski spacetime, to
be subm. for publ.}

{\footnotesize Hestenes, D. (1966). \emph{Space-time Algebra}, Gordon\&Breach,
New York.}

{\footnotesize Hestenes, D. and Sobczyk, G. (1984). \emph{Clifford Algebra to
Geometrical Calculu}s, Reidel, Dordrecht.}

{\footnotesize Fern\'{a}ndez, V. V., Moya, A. M., and Rodrigues, W. A., Jr.
(2000a). Covariant derivatives on Minkowski manifolds, in Ryan, J. and
Sproessig, W. (eds.), \emph{Proc. 5th Int.} \emph{Conf. on Clifford Algebras
and their Applications in Mathematical Physics}, vol 1: \emph{Algebra and
Physics}, pp. 367-392, Birkhauser, Boston, 2000.}

{\footnotesize Fern\'{a}ndez, V. V., Moya, A. M., and Rodrigues, W. A., Jr.
(2000b). The algebraic theory of connections, differential and Lie operators
for Clifford and extensor fields, to be subm. for publ. (2003)}

{\footnotesize Lasenby, A., Doran, C., and Gull, S. (1993). A Multivector
Derivative Approach to Lagrangian Field Theory, \emph{Found. Phys}.
\textbf{23}, 1329-1356.}

{\footnotesize Lasenby, A., Doran, C., and Gull, S. (1998). Gravity, gauge
theories and geometric algebra, \emph{Phil. Trans. R. Soc}. \textbf{356},
487-582.}

{\footnotesize Lounesto, P. (1997). \emph{Clifford Algebras and Spinors.}
Cambridge University Press, Cambridge.}

{\footnotesize Rodrigues, W. A., Jr., and Rosa, M. A. F. (1989). The meaning
of time in relativity and Einstein later view of the twin paradox,
\emph{Found. Phys}.\textbf{19}, 705-724.}

{\footnotesize Rodrigues, W. A., Jr., de Souza, Q. A. G. (1993). The Cifford
bundle and the nature of the gravitational field, \emph{Found. Phys.}
\textbf{23}, 1456-1490.}

{\footnotesize Rodrigues, W. A., Jr., de Souza, Q. A. G., and Vaz, J., Jr.
(1994). Lagrangian formulation in the Clifford bundle of the Dirac-Hestens
equation on a Riemann-Cartan manifold, in \emph{Gravitation}: \emph{The
Spacetime Structure}. Proc. SILARG VIII, \'{A}guas de Lind\'{o}ia, SP, Brazil,
25-30, July, 1993, P. Letelier and W. A. Rodrigues, Jr., eds., World
Scientific, Singapore, pp. 522-531.}

{\footnotesize Sachs, R. K. and Wu, H. (1977). \emph{General Relativity for
Mathematicians}, Springer-Verlag, New York.}

{\footnotesize Rodrigues, W. A., Jr., de Souza, Q. A. G, Vaz, J., Jr.and
Lounesto, P. (1995). Dirac-Hestenes Spinor Fields on Riemann-Cartan Manifolds,
\emph{Int. J. Theor. Phys.} \textbf{35}, 1849-1900.}

\end{document}